\documentclass[aps,prb,twocolumn,groupedaddress]{revtex4}
\usepackage{graphics,epsfig,comment}
\begin{document}
\title{Room Temperature Superconductivity Revolution: 
\\ Foreshadowed by Victorians,
  Enabled by Millenials}
\author{Warren E. Pickett}
\affiliation{Department of Physics, University of California Davis,
   Davis CA 95616}

\date{\today}
\begin{abstract}
Room temperature superconductivity has been the most prominent, highly ambitious, 
but still imaginable, acme of materials physics for half a century. The struggle
toward this revolution was foreshadowed by a Victorian novelist and championed,
unsuccessfully, by dogged physicists in the 1960s to 1980s who had a workable
theory but uncompliant materials. Discovery of superconductivity of H$_3$S at
200 K in the 160-200 GPa pressure range has renewed anticipation of yet higher
values of the critical temperature T$_c$. With the several reports of metalization
of hydrogen, and theoretical extensions enabled by modern algorithms and unprecedented
computational hardware and spurred forward by the Materials Genome Initiative, 
it is possible that the room temperature 
precipice has thereby already been breached in a silent revolution. 
This concise note draws analogies of this development with an earlier revolution. 
\end{abstract}
\maketitle

~
\vskip 5mm
\begin{center} {\bf Chapter 1. Drawing the lines of battle} \end{center}

{\it It was the best of times; it was the worst of times.}\cite{Dickens}
The quest for a 
high temperature superconductivity (HTS) revolution roiled in the 1960s, 
instigated by Bernd Matthias who was also its most zealous experimental practitioner. 
The maximum critical temperature T$_c^{max}$ increased modestly from 17K to 23K 
from 1955 to 1973. In that year Bruce Friday, in a letter to Physics 
Today,\cite{Friday} recognized a trend in T$_c^{max}$ from the 
discovery of superconductivity to that time, 
and endeavored to sooth the hopeful platoons of materials researchers with 
analysis that indicated the linear-in-time increase, reproduced in Fig. 1, 
extrapolated to room temperature 
superconductivity around year 2840. No revolutions would be necessary, nor were any 
envisioned, by Friday or by the disillusioned proletariat. Friday's observation served 
also as an example of quantum observation: that moment in 1973 signaled the collapse from a 
state of steady increase in T$_c^{max}$ to a state of no increase in conventional 
T$_c^{max}$ for three decades.\cite{Fullerides} 

\begin{figure}[!ht]
\includegraphics[width=0.65\columnwidth]{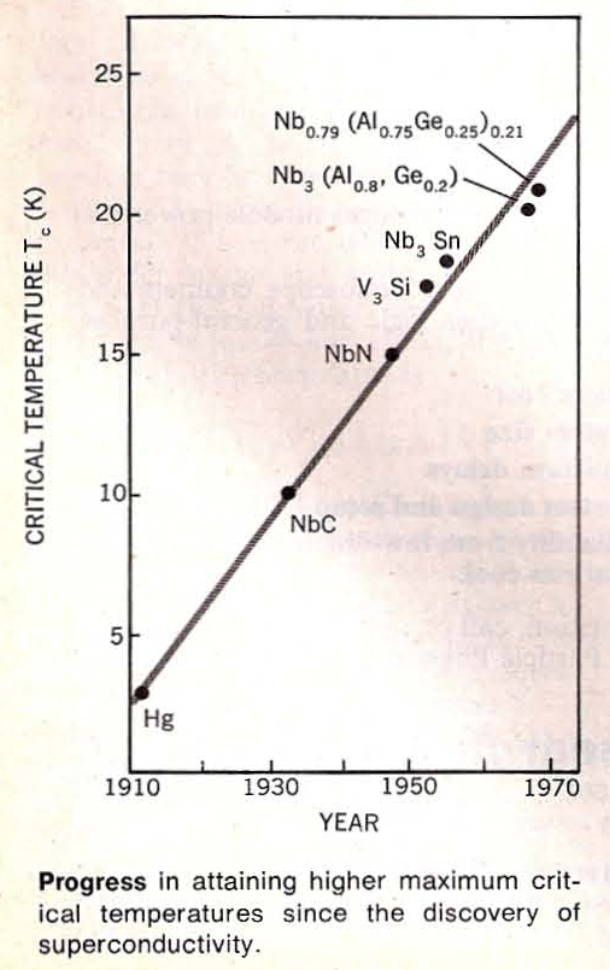}
\caption{
Bruce Friday's plot of maximum T$_c$ versus time, from the discovery of
superconductivity in 1911 to the time of this plot in 1973.
Extrapolation of his linear fit indicated room temperature superconductivity
in the year 2840.
}
\label{Tc-vs-time}
\end{figure}

During the 1960s Matthias formalized\cite{Matthias} his 
royal decree for higher T$_c$: (1) use transition metal (TM) based materials, 
(2) specific electron/atom ratios are best, and (3) cubic symmetry is preferred. 
These rules were formulated from elemental TMs, TM carbides and nitrides, and a 
few other TM-dominant compounds, viz. Nb$_3$Sn, whose simple A15 structure is pictured
in Fig. 2. The gauntlet was 
laid down to theorists for successful predictions, with notoriety and careers
as the prize, or more likely the cost.  Theorists responded with the 
premier advancements of that
age. Scalapino and coworkers formulated\cite{Scalapino} 
the implementation of Migdal-Eliashberg
pairing theory of electron-phonon coupling in a material-specific manner.  
Phil Allen and Bob Dynes 
demonstrated that M-E theory was correct and robust,\cite{Allen} 
and that the theoretical foundation
imposed no limit on T$_c$. Material systems that seemed poised to confront the 
frontier of strong coupling and much higher T$_c$
led only to structural instabilities, or to competing order such as magnetic
or charge- and spin-density waves, that lay beyond the quantitative theory of the day.
{\it It was the age of wisdom, it was the age of foolishness.}

\begin{figure}[!ht]
\includegraphics[width=0.65\columnwidth]{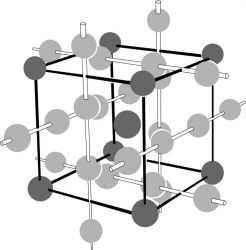}
\caption{
The simple, cubic, and beautiful crystal structure of the
early class of A15
highest temperature superconductors, viz.Nb$_3$Sn. Dark spheres: Sn;
gray spheres: Nb. A dominant feature of the  structure is the chain
of Nb atoms directed along each of the cubic axes.
}
\label{A15-structure}
\end{figure}

\begin{center}{\bf Chapter 2. Theoretical uprising} \end{center}

A rigorous formalism and computational basis for electronic structure emerged -- 
density functional theory [DFT] -- whereby the microscopic understanding and 
quantitative prediction of electronic structures ballooned in the ‘70s and ‘80s. 
By 1980 the Matthias rules had been understood on a quantitative basis: 
(i) TMs promote directional bonding -- covalent but still metallic -- which promotes 
stiff lattices and hefty electron-phonon scattering matrix elements, 
(ii) in each of these classes, the density of states at the Fermi level N(0) peaks 
at specific electron/atom values because each class is rigid band like, 
(iii) cubic materials have no lattice constant ratio $b/a$ or $c/a$ 
that can be relaxed to relieve 
electronic stress [high N(0)], promoting the superconducting onset but competing 
instabilities as well. The theoretical musketeers were demonstrating strength, 
portending closer interplay between experiment and theory as driving the next 
revolution.  {\it It was the spring of hope, it was the winter of despair}:
establishing a detailed theoretical foundation was not leading to discovery of 
better superconductors.

In 1980 the computational artillery was not in a position to predict T$_c$ from 
first principles; the advance of formal theory had not produced the best of all possible 
worlds. Phonon frequencies had to be taken from experiment, matrix elements from 
potentials of rigidly moving atoms; moreover, promising predictions tended to 
produce unstable materials. 
But {\it there is prodigious strength in sorrow and despair.}
By the 1990s, phonons could be computed 
accurately from first principles and matrix elements calculated from rigorous 
linear response, solving the problem except for the relatively small but curious 
Coulomb repulsion. The theoretical struggle was decided by the first decade of the 
21st century, with Hardy Gross's formulation of and implementation of DFT for 
superconductors.\cite{Gross} For materials with weakly interacting electrons but including 
strongly coupled electrons and phonons, calculations of T$_c$ became accurate to 
perhaps 5\% (sometimes claimed to be better than that).  {\it It was the epoch of 
belief, it was the epoch of credulity.}

The HTS revolution of copper-based oxides in 1986, extending to 160K under pressure, 
and iron-based materials at 55K in 2008 extending to 75K in single layers, hugely 
energized the proletariat. The superconducting but unyielding A15 structure 
royalty had been 
magnificently overthrown; commoners could synthesize high T$_c$ samples.  After 
seeming eons (30 years, and 8 years, respectively) theory has produced no 
quantitative picture of pairing in these materials. 

\begin{center} {\bf Chapter 3. Unleashing the computational artillery}\end{center}

The Materials Genome Initiative 
(MGI) of 2011 formalized a new paradigm: introduce large scale, high throughput 
computation into the synthesis \& characterization cycle to accelerate the design 
\& discovery of novel materials with improved functionality. To date, however, 
applying the MGI approach to superconductors has been limited to a few intrepid 
musketeers. Mathias Klintenberg and Olle Eriksson\cite{Klintenberg} 
searched for cuprate-like electronic 
structures using modern battlefield technology: high-throughput computing and 
data-filtering algorithms. With a similar goal, the EFRC Center for Emergent 
Superconductivity has focused on searches based on structural motifs. Success is 
yet to be demonstrated, but such challenges are meant to be confronted and overcome. 
{\it It was the season of light, yet it remained the season of darkness.}

Thus on the topic of superconducting materials, as of 2015 the MGI approach had yet 
to knit new materials to higher superconducting T$_c$; {\it the spring of 
hope remained hidden beyond the winter of despair.} Still, it may now not be too 
soon to reconsider how MGI and the available computing capabilities can best be 
applied, and Mike Norman has provided a broad overview of MGI in relation to the 
search for better superconducting materials.\cite{Norman}  The more realistic 
promise for HTS+MGI remains, at this writing, within the phonon-coupled paradigm, 
though early efforts\cite{Klintenberg} had focused on the cuprate paradigm. 

Not only was the MgB$_2$ insurgency of 2001, led by 
Jun Akimitsu's group,\cite{MgB2} a stunning 
overthrow of the established order, so also was the rapid response with which DFT 
musketeers devined the underlying mechanism –- covalent bonds driven metallic by 
chemistry –- and reproduced the observed T$_c$=40K, remarkable for a phonon mechanism. 
Though phonon mediated and far from optimal,\cite{Pickett} 
MgB$_2$ violated each of the emperor’s 
(Matthias's) dictates. Transition metals are not essential and not even optimal, 
being overthrown by a 
broader edict: covalent bonding in a metal. Large N(0) is not the target, high 
T$_c$ is the target. Hexagonal and two dimensional can be better than cubic. 
Very simple to understand, very difficult to improve on, as a handful of attempts 
has demonstrated.  {\it Were we all going directly to heaven, or were we all 
going direct the other way?}

\begin{center}{\bf Chapter 4. A synergistic revolution} \end{center}

The revolutionary announcement in 2015 by Mikhail Eremets' group\cite{Eremets} 
of superconductivity at 200K under very high pressure (160-200 GPa), in 
putrid but otherwise unremarkable hydrogen sulfide, provided a primitive 
application of the MGI paradigm, in that the extraordinary value of T$_c$ 
resulted from theory spurring experiment and further theory, rather than the
standard paradigm of experiment spurs theory. 

\begin{figure}[!ht]
\includegraphics[width=0.65\columnwidth]{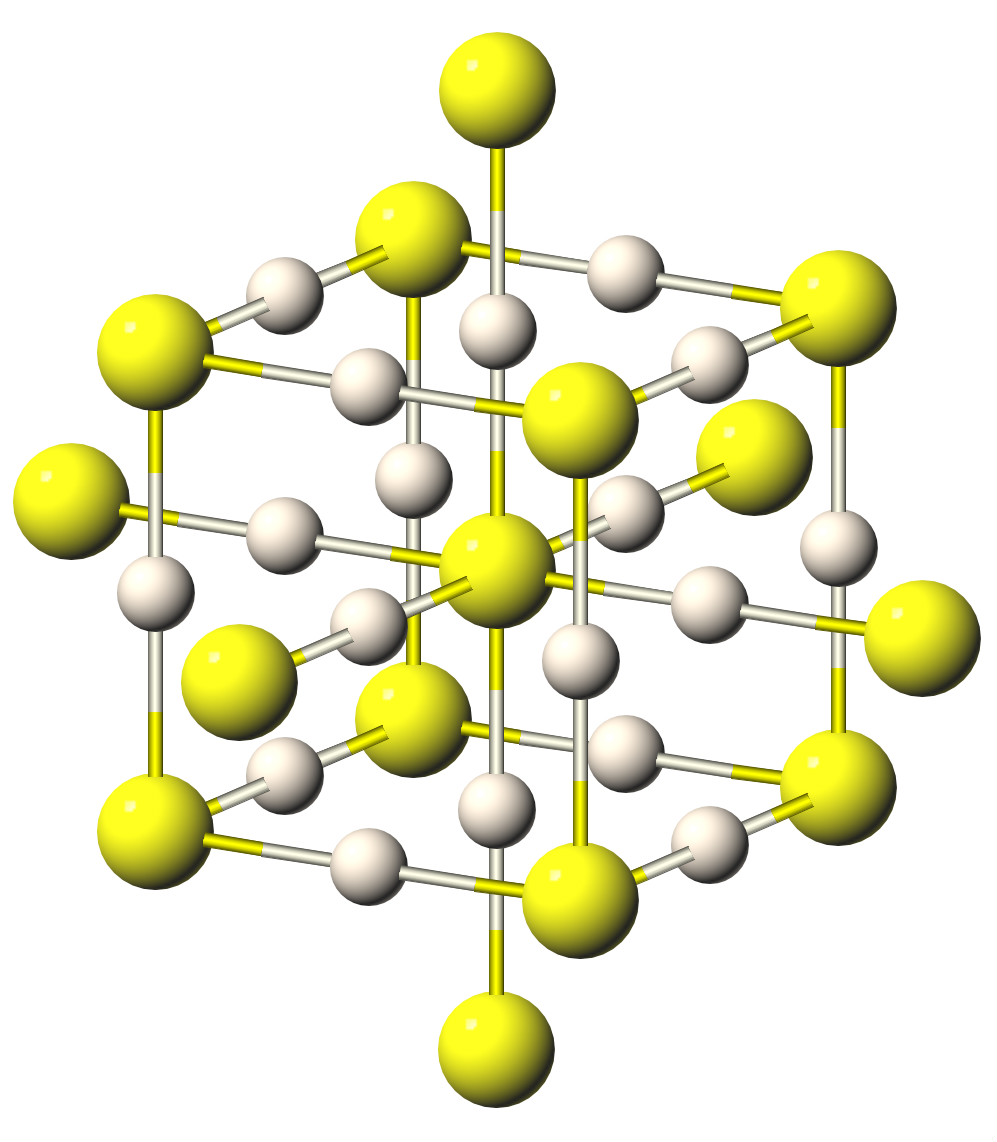}
\caption{
The simple, beautiful, and cubic bcc crystal structure of the current
highest temperature superconductor H$_3$S. Yellow spheres: sulfur;
gray spheres: hydrogen. Comparisons with the A15 structure
are striking: each is one of the three simplest cubic $A_3B$ 
structures; each has dominant
electron-phonon coupling arising from one atomic species; 
each displays a sharp and narrow density
of states peak lying precisely at the Fermi level. 
}
\label{Structure}
\end{figure}

The earlier computational study of H$_2$S by Yanming Ma's group\cite{Ma} predicting 
T$_c$$\sim$80K under pressure stimulated the experimental effort of Eremets, 
which synergistically spurred Tian Cui's group to extend the predictions\cite{Cui} to 
H$_3$S. Their revolutionary effort predicted the outrageous value of T$_c\sim$ 200 K 
at extremely high pressure in the 200 GPa (two million atmospheres) regime.  Eremets' 
confirmation of this prediction demonstrated the remarkable power of theory, 
first to identify bcc H$_3$S as the stable phase from a variety of competing structures
(an MGI-inspired approach), 
and finally to predict unbelievably high T$_c$ correctly. 

Remarkably, 
the experimental discovery publication\cite{Eremets} references seven theory 
papers explaining, and agreeing on, the mechanism and the very high 
T$_c$. This experiment-theory inversion was enabled by the posting of a preprint
that had been arXiv'd months earlier.\cite{arXiv} 
The strong theoretical agreement provides
the broad view: DFT-based Migdal-Eliashberg theory is robust at least up into
the room temperature regime. Deeper analysis is more arresting, with further 
questions emerging: what is the impact of the two van Hove singularities that 
conspire to put the Fermi level in the best possible position for large N(0) 
but in an extremely narrow peak? is anharmonicity good or bad for T$_c$? how 
much does the quantum nature of the proton affect the properties of H$_3$S, 
particularly the isotope shift of T$_c$? These unsettling loose ends are succumbing to
modern theory and computation. 

\begin{center}{\bf Chapter 5. Visualizing utopia} \end{center}

As happens after a revolution, new and compelling issues emerge: 
can room temperature superconductivity be achieved? can related (possibly 
metastable) materials be tormented into a very high T$_c$ phase at much 
reduced pressure?  {\it We have everything before us, or have we nothing 
before us.}

The search for HTS in hydrogen-based materials owes much to the vision and 
persistence of Neil Ashcroft,\cite{Ashcroft} yet the 
success in hydrogen sulfide just mentioned 
instills pessimism: have we perhaps gone from having too much to work on to 
having nothing left to accomplish? Stepping into this saga personally, we here 
boldly propose that the formidable battlement sheltering the 
holy grail has been 
breached: a room temperature superconducting phase has recently been achieved, 
though yet undetected due to the challenges of making the necessary measurements 
at ultrahigh pressure. Several reports of metalization of hydrogen in the range of 
400-500 GPa have appeared, most vociferously by Ike Silvera’s brigade\cite{Silvera} 
but earlier by other groups,\cite{Others} 
although the data have not convinced everyone on the battlefield. 

Clearly modern electronic structure theory is confronted with a 
huge challenge in this regime. It seems however that again, as for H$_3$S, this 
gauntlet has already been challenged and overcome. Several groups have 
contributed to the determination of the hydrogen phase diagram at ultrahigh 
pressures including the quantum nature of the proton, which has a tangible influence. 
This quantum uncertainty 
affects the structure-pressure-temperature phase diagram but might have less 
affect on the electron-phonon coupling strength $\lambda$. Ceperley's group 
had carried out the necessary calculations\cite{Ceperley} in the predicted crystal structure 
and found T$_c$ to be at or above 350K at pressures attained so far,
with substantially higher critical temperatures predicted at increased pressure. 
There can be no argument that room temperature superconductivity has provided 
the acme of superconductivity aspirations, and it quite plausibly has been 
achieved. It remains for experimentalists to confront the challenge: make 
reproducible measurements to test the predictions.

\begin{center}{\bf Chapter 6. The next uprising}\end{center}

With verification of this proposal, viz. that room temperature superconductivity 
has been achieved, the best of times would seem to be within sight. The next 
challenge is in place: to produce these elevated critical temperatures at much 
reduced pressure. The theoretical prowess has been verified, and it can be 
reasonably expected to accelerate the path, perhaps leading the way, toward 
meeting future challenges. While the MGI is a far more broadly based initiative 
than superconductivity, this high visibility field provides 
an example of much needed success 
that should serve as an inspiration to the many researchers who are engrossed 
in this new crusade. This is no time for the computational musketeers to take 
the conservative path, which could be stated: {\it keep where you are because, 
if (one) should make a mistake, it could never be set right in your lifetime.}
 Boldness is de regueur.

~
\vskip 3mm
The MGI concept, and its implementation, is still evolving as it advances. 
Major discoveries are yet to appear. Still, several groups may be able to say 
about their effort at design \& discovery of high T$_c$: {\it it is a far, 
far better thing that I do, than I have ever done; it is a far, far better 
rest to go to than I have ever known.}

Acknowledgments. 
The author's involvement in this area of research has over the years
been supported in various periods by NSF, DOE, and ONR.


\begin{thebibliography}{99}
\bibitem{Fullerides} The fullerides and (Ba,K)BiO$_3$ involve the lattice
  but not in prosaic form.
\end{thebibliography}
\end{document}